\documentclass[%
a4paper,       
UKenglish,     
DIV10,          
useAMS,usenatbib,referee, doublespace]%
{scrartcl}\usepackage[]{graphicx}\usepackage[]{color}
\makeatletter
\def\maxwidth{ %
  \ifdim\Gin@nat@width>\linewidth
    \linewidth
  \else
    \Gin@nat@width
  \fi
}
\makeatother

\definecolor{fgcolor}{rgb}{0.345, 0.345, 0.345}

\usepackage{framed}
\makeatletter
 {\par\unskip\endMakeFramed%
 \at@end@of@kframe}
\makeatother

\definecolor{shadecolor}{rgb}{.97, .97, .97}
\definecolor{messagecolor}{rgb}{0, 0, 0}
\definecolor{warningcolor}{rgb}{1, 0, 1}
\definecolor{errorcolor}{rgb}{1, 0, 0}
\newenvironment{knitrout}{}{} 

\usepackage{alltt}

\usepackage[utf8]{inputenc}
\usepackage[UKenglish]{babel}
\usepackage[T1]{fontenc}
\usepackage{textcomp}
\usepackage{verbatim}
\usepackage[normalem]{ulem}
\usepackage{setspace}
\usepackage{hyperref}  
\usepackage{url}                
\usepackage[sumlimits, intlimits, namelimits]{amsmath} 
\usepackage{amssymb}            
\usepackage[caption = true]{subfig} 
\usepackage{array}              
\usepackage{booktabs}           
\usepackage{ae}                 
\usepackage[longnamesfirst,round]{natbib} 
\usepackage{color}              
\usepackage[sc]{mathpazo}       
\usepackage{helvet}             
\usepackage{todonotes}          
\newcommand{\hl}{\textcolor{black}}

\def \blinded {1}  

\hypersetup{
  bookmarksopen=true, 
  breaklinks=true,
  pdftitle={The harmonic mean chi-squared test}, 
  pdfsubject={Statistics},
  pdfkeywords={Statistics, Evidence},
  colorlinks=true,
  linkcolor=blue,
  anchorcolor=black,
  citecolor=teal,
  urlcolor=purple
}

\usepackage{xifthen}            
\usepackage{array}
\usepackage{amssymb}

\newcommand{\blanco}[1]{  } 

\newcommand{\latin}[1]{\textit{#1}}

\newcommand{\abk}[1]{\mbox{#1}\xdot}
\DeclareRobustCommand\xdot{\futurelet\token\Xdot}
\def\Xdot{%
  \ifx\token\bgroup.%
  \else\ifx\token\egroup.%
  \else\ifx\token\/.%
  \else\ifx\token\ .%
  \else\ifx\token!.%
  \else\ifx\token,.%
  \else\ifx\token:.%
  \else\ifx\token;.%
  \else\ifx\token?.%
  \else\ifx\token/.%
  \else\ifx\token'.%
  \else\ifx\token).%
  \else\ifx\token-.%
  \else\ifx\token+.%
  \else\ifx\token~.%
  \else\ifx\token.%
  \else.\ %
  \fi\fi\fi\fi\fi\fi\fi\fi\fi\fi\fi\fi\fi\fi\fi\fi%
}

\newcommand{\eg}{\abk{\latin{e.\,g}}}
\newcommand{\ie}{\abk{\latin{i.\,e}}}
\newcommand{\cf}{\abk{\latin{cf}}}

\newlength{\halbebreite}
\DeclareMathOperator{\Nor}{N} 
\DeclareMathOperator{\Levy}{Levy} 
  
\DeclareMathOperator{\Ga}{G} 
\DeclareMathOperator{\IG}{IG} 

\renewcommand{\P}{\operatorname{\mathsf{Pr}}} 

\newcommand{\partialv}[3][1]{%
\ifthenelse{#1 = 1}{\frac{\partial #2}{\partial #3}}{\frac{\partial^{#1} #2}{\partial #3^{#1}}}
} 

\newcommand{\partials}[3][1]{%
\ifthenelse{#1 = 1}{\frac{d #2}{d #3}}{\frac{d^{#1} #2}{d #3^{#1}}}
} 

\newcommand{\dseps}[2][1]{%
\ifthenelse{#1 = 1}{\frac{d}{d #2}}{\frac{d^{#1}}{d #2^{#1}}}
}

\newcommand{\dsepv}[2][1]{%
\ifthenelse{#1 = 1}{\frac{\partial}{\partial #2}}{\frac{\partial^{#1}}{\partial #2^{#1}}}
}

\newcommand{\ml}[2][1]{
\ifthenelse{#1 = 1}%
 {\hat{#2}_{\scriptscriptstyle{\mathrm{ML}}}}%
 {\hat{#2}^{#1}_{\scriptscriptstyle{\mathrm{ML}}}}
}

\newcommand{\zTwoMin}{{z_{\mbox{\scriptsize min}}^2}}

\newcommand\reduline{\bgroup\markoverwith
{\textcolor{red}{\rule[-0.5ex]{2pt}{0.4pt}}}\ULon}

\IfFileExists{upquote.sty}{\usepackage{upquote}}{}
\IfFileExists{upquote.sty}{\usepackage{upquote}}{}
\begin{document}

\title{The harmonic mean $\chi^2$ test \\ to substantiate scientific findings}
  \ifcase\blinded 
\author{}
\or          
\author{Leonhard
  Held\\ 
  Epidemiology, Biostatistics
  and Prevention Institute (EBPI)\\ 
  and Center for Reproducible Science (CRS) \\
  University of Zurich\\ Hirschengraben 84,
  8001 Zurich, Switzerland\\ Email: \texttt{leonhard.held@uzh.ch}}
\fi

\maketitle

\begin{center}
\begin{minipage}{12cm}
  \textbf{Abstract}: Statistical methodology plays a crucial role in
  drug regulation. Decisions by the FDA or EMA are typically made
  based on multiple primary studies testing the same medical product,
  where the two-trials rule is the standard requirement, despite a
  number of shortcomings.  A new approach is proposed for this task
  based on the (weighted) harmonic mean of the squared study-specific
  test statistics.  Appropriate scaling ensures that, for any number
  of \hl{independent} studies, the null distribution is a
  $\chi^2$-distribution with one degree of freedom.  \hl{This gives
    rise to a new method for combining one-sided $p$-values and
    calculating confidence intervals for the overall treatment effect.}  Further
  properties are discussed and a comparison with the two-trials rule
  is made, as well as with alternative research synthesis methods.  An
  attractive feature of the new approach is that a claim of success
  requires each study to be convincing on its own to a certain degree
  depending on the \hl{overall} significance level and the number of studies.  
  The new approach is motivated by and applied to data from five 
  clinical trials investigating the effect of
  Carvedilol for the treatment of patients with moderate to severe
  heart failure.
\end{minipage}
\end{center}


\section{Introduction}
Research synthesis has been characterized as the process of combining
the results of multiple primary studies aimed at testing the same
conceptual hypothesis. Meta-analysis is the preferred technique of
quantitative research synthesis, as it provides overall effect
estimates with confidence intervals and $p$-values through pooling of study results
and
allows for the incorporation of heterogeneity between studies.
However, meta-analysis can be criticized as a too weak technique if
the goal is to substantiate an original claim through one or more
additional independent studies. Specifically, a significant overall 
effect estimate may occur even if some of the individual studies have
not been convincing on its own, perhaps even with effect estimates in
the wrong direction.  This may be acceptable if the unconvincing
studies have been small, but seems less tolerable if each study was
well-powered and well-conducted.

For example, consider the results from 5 clinical trials on the effect
of Carvedilol, a beta- and alpha-blocker and an antioxidant drug for
the treatment of patients with moderate to severe heart failure, 
on mortality \citep[\cf][Table 1]{Fisher1999b}.  One-sided
$p$-values (from log-rank tests) and estimated hazard ratios (HR) 
are shown in Table \ref{tab:tab3}, 
indicating a reduction in mortality
between 28 and 78\% across the different
studies.

\begin{table}[ht]
\centering
\begin{tabular}{rrrrr}
  \hline
study number & $p$-value & HR & log HR & SE \\ 
  \hline
220 & 0.00025 & 0.27 & -1.31 & 0.41 \\ 
  240 & 0.0245 & 0.22 & -1.51 & 0.85 \\ 
  223 & 0.128 & 0.72 & -0.33 & 0.29 \\ 
  221 & 0.1305 & 0.57 & -0.56 & 0.51 \\ 
  239 & 0.2575 & 0.53 & -0.63 & 1.02 \\ 
   \hline
\end{tabular}
\caption{Results from 5 clinical trials on the effect of Carvedilol for the treatment of
patients with moderate to severe heart failure. Shown are one-sided $p$-values, \hl{estimated} hazard ratios (HR), and the associated log hazard ratios (log HR) with standard errors (SE).} 
\label{tab:tab3}
\end{table}

A meta-analysis could be applied to the data shown in Table
\ref{tab:tab3}, but the drug regulation industry (including the
U.S.~Food and Drug Administration, or FDA) typically relies instead on
the "two-trials rule" \citep{senn:2007,kay:2015}, also known as the
``two pivotal study paradigm'' \citep{pmid26753552}, for approval.
This simple decision rule requires ``at least two adequate and
well-controlled studies, each convincing on its own, to establish
effectiveness'' \citep[p.~3]{FDA1998}.  This is usually achieved by
independently replicating the result of a first study in a second
study, both significant at one-sided level $\alpha=0.025$. However, in
modern drug development often more than two trials are conducted and
it is unclear how to extend the two-trials rule to this
setting. Requiring at least 2 out of $n>2$ studies to be significant
is too lax a criterion if the results from the non-significant studies
are not taken into account at all.  On the other hand, requiring all
$n$ studies to be significant is too stringent. This problem applies
to the Carvedilol example, where two trials are significant at the
2.5\% level (one just with $p=0.0245$) but where it is
unclear whether the remaining three studies (with $p$-values
0.128, 0.1305 and 0.2575)
can be considered as sufficiently ``convincing on its own.''  

This has led statistical researchers to discuss the possibility of
pooling the results from the different studies into one $p$-value
\citep{Fisher1999, DarkenHo2004, Shun_etal2005}.  Fisher's
method to combine $p$-values \citep{Fisher-1958} is often used for
this task, \eg in \cite{Fisher1999b} for the Carvedilol
example. However, Fisher's method shares the problems of a
meta-analysis as it can produce a significant overall result even if
one of the trials was negative. For example, one completely
unconvincing trial with (one-sided) $p=0.5$ combined with a convincing
second one with $p=0.0001$ would give Fisher's
$p=0.0005 < 0.000625=0.025^2$, so a claim of success with respect to
the Type I error rate of the two-trials rule.  On the other hand, two
trials both with $p=0.01$ would give Fisher's $p=0.001$, so would
not be considered as successful. Both decisions seem undesirable from a regulator's
perspective.

The two-trials rule therefore remains the standard in drug regulation,
but has additional deficiencies even for $n=2$ studies, where
independent $p$-value thresholding at $0.025$ may lead to decisions
that are the opposite to what the evidence warrants. For example, two
trials both with $p=0.024$ will lead to drug approval but carry less
evidence for a treatment effect than one trial with $p=0.026$ and the
other one with $p=0.001$, which would, however, not pass the
two-trials rule.  \cite{Rosenkrantz2002} has therefore proposed a
method to claim efficacy if one of two trials is significant while the
other just shows a trend. He combines the two-trials rule with
Fisher's method and a relaxed criterion for significance of the two
individual trials, say $2 \alpha$. A similar approach has been
proposed by \citet{Maca2002} using Stouffer's pooled rather than
Fisher's combined method. The arbitrariness in the choice of the
relaxed significance criterion is less attractive, though, and it is
not obvious how to extend the methods to results from more than two
studies.

In this paper I develop a new method that addresses these
issues and leads to more appropriate inferences, the harmonic mean
$\chi^2$ test described in Section \ref{sec:sec2}.  At the Type I error rate 
$0.025^2$ of the two-trials rule, the proposed test
comes to opposite conclusions for the examples mentioned
above: In contrary to Fisher's method, it leads to approval of two
trial both with $p=0.01$, but not to approval if one has $p=0.0001$ and the other
one $p=0.5$. Contrary to the two-trials rule, it leads to
approval of one trial with $p=0.026$ and the other one with $p=0.001$,
but not to approval if both trials have $p=0.024$. The work is
motivated from a recent proposal how to evaluate the success of
replication studies \citep{held2020} and is based on the harmonic mean
of the squared $Z$-scores. It can include weights for the individual
studies and can be calibrated to ensure exact Type I error
control \hl{and to compute an overall $p$-value, see Section \ref{sec:Pvalues}}. Furthermore, the new approach
implies useful bounds on the study-specific $p$-values, thus formalizing the meaning of ``at least two
adequate and well-controlled studies, {\em each convincing on its
  own}''. \hl{It can also be used to calculate a confidence interval for the
  overall treatment effect, see Section \ref{sec:CI}.} 
The approach will be compared to the
two-trials rule in Section \ref{sec:sec3} and applied to the Carvedilol data in Section
\ref{sec:sec.app}. I close
with some discussion in Section \ref{sec:sec5}.  

\section{The harmonic mean $\chi^2$ test}
\label{sec:sec2}

Suppose one-sided $p$-values $p_1, \ldots, p_n$ are available from $n$
independent studies. How can we combine the $p$-values into one
$p$-value? \citet{cousins2007annotated} compares some of the more
prominent papers on this topic. Among them is Stouffer's method, which
is based on the $Z$-scores $Z_i = \Phi^{-1}(1-p_i)$, here
$\Phi^{-1}(.)$ denotes the quantile function of the standard normal
distribution.  Under the assumption of no effect, the test statistic
$Z = \sum_{i=1}^n Z_i / \sqrt{n}$ is standard normally
distributed. The corresponding $p$-value forms the basis of the
``pooled-trials rule'' and is equivalent to investigate significance
of the overall effect estimate from a fixed-effects meta-analysis
\citep[Section 12.2.8]{senn:2007}. 
Fisher's method is also commonly used and compares
$-2 \sum_{i=1}^n \log p_i$ with a $\chi^2$-distribution with $2 n$
degrees of freedom to compute a combined $p$-value.  
Both Stouffer's and Fisher's method can be extended
to incorporate weights, where the null distribution of the latter
does no longer have a convenient form \citep{Good1955}.
\hl{There is a large
literature on the comparison of these and other methods for the combination
of $p$-values, such as \citet{LittellFolks1973,BerkCohen1979,Westberg1985,HeardRubin-Dlanchy2018}.}

Here I propose a \hl{new approach} to assess the overall evidence for a treatment effect
based on the 
harmonic mean $Z_H^2 = n/{\sum_{i=1}^n 1/Z_i^2}$ of the squared $Z$-scores:
\begin{equation}\label{eq:eq0}
  X^2 = n \, Z_H^2 = \frac{n^2}{\sum\limits_{i=1}^n 1/Z_i^2}.
\end{equation}
This form is motivated from the special case of $n=2$ successive
studies, one original and one replication, where a reverse-Bayes approach for the assessment of
replication success has
recently been described \citep{held2020}. If the two studies have equal precision (\ie sample size), the
assessment of replication success does not depend on the order of the
two studies and is based on the test statistic $1/(1/Z_1^2 + 1/Z_2^2)$,
compare \citet[equation (9)]{held2020}. Equation \eqref{eq:eq0}
extends this to $n$ studies with an additional multiplicative factor
$n^2$, which ensures that the null distribution of \eqref{eq:eq0} does not
depend on $n$.  
Weights $w_1, \ldots, w_n$ can also be introduced in \eqref{eq:eq0},
then the test statistic
\begin{equation}\label{eq:eq1}
  X_w^2 =  \frac{w^2} {\sum\limits_{i=1}^n w_i/Z_i^2} \mbox{ where $w = \sum\limits_{i=1}^n \sqrt{w_i}$}
\end{equation}
should be used. The factor $w^2$
ensures that the null distribution of \eqref{eq:eq1} does not depend on the weights $w_1, \ldots, w_n$ nor on $n$. 

The specific form of \eqref{eq:eq1} deserves some
additional comments. In practice we often have
$Z_i = \hat \theta_i/\sigma_i$ where $\sigma_i = \kappa/\sqrt{m_i}$ is the standard error
of the effect estimate $\hat \theta_i$, $\kappa^2$ is the
one-unit variance and $m_i$ the sample size of study $i$. 
If we use weights
$w_i= 1/\sigma_i^2$ equal to the precision of the effect estimates, \eqref{eq:eq1} can be written as
the unweighted harmonic mean $\hat \theta_H^2$ of the squared effect estimates $\hat \theta_i^2$ times a scaling factor $w^2/n$: 
\begin{equation}\label{eq:eq3}
  X^2_w = {w^2}/{n} \cdot \hat \theta_H^2 \mbox{ where } w=\sum\limits_{i=1}^n \sqrt{m_i}.
\end{equation}
In the special case of equal sample sizes  $m_1=\ldots m_n=m$, the scaling factor
reduces to $n\, m$. 

There is a subtle difference between the two formulations
$\eqref{eq:eq0}$ and $\eqref{eq:eq3}$.  The unweighted test statistic
\eqref{eq:eq0} is based on the harmonic mean of the squared study-specific
test statistics $Z_i^2$, $i=1,\ldots,n$.  If we increase the sample size
of the different studies, \eqref{eq:eq0} will therefore also tend to increase if
there is a true non-zero effect. However, the test statistic
\eqref{eq:eq3} is based on the harmonic mean $\hat \theta_H^2$ of the squared study-specific
effect estimates $\hat \theta_i^2$, 
which should
not be much affected by any increase of study-specific sample sizes because the
study-specific estimates $\hat \theta_i$ should then stabilize around their true values. 
It is the scaling
factor $w^2/n$ that will react to an increase in study-specific
sample sizes.  The test statistic $\eqref{eq:eq3}$ can thus be
factorized into a component depending on sample sizes and a
component depending on effect sizes. 

\subsection{\hl{$P$-values}}
\label{sec:Pvalues}

Using properties of L\'evy distributions it can be shown that under
the null hypothesis of no effect, the distribution of both
\eqref{eq:eq0} and \eqref{eq:eq1} is $\chi^2$ with one degree of
freedom, see Appendix \ref{app:app1} for details.  We can thus compute
an overall $p$-value $p_H$ from \eqref{eq:eq0} or \eqref{eq:eq1} based
on the $\chi^2(1)$ distribution function.  However, we have to be
careful since \eqref{eq:eq0} does not take the direction of the
effects into account. Usually we are interested in a pre-defined
direction of the underlying effect, say $H_1$: $\theta > 0$ against
$H_0$: $\theta=0$ and we will have to adjust for the fact that
\eqref{eq:eq0} and \eqref{eq:eq1} can be large for any of the $2^n$
possible combinations of the signs of $Z_1, \ldots, Z_n$, with all
these combinations being equally likely under the null
hypothesis. Since we are interested only in the case where all signs
are positive, we have to adjust the $p$-value accordingly.

To be specific, suppose all studies have a positive effect and the observed test
statistic \eqref{eq:eq0} or \eqref{eq:eq1} is $X^2=x^2$, respectively $X^2_w=x^2$ and let
$x = + \sqrt{x^2}$. The overall
$p$-value from the proposed significance test is then 
\begin{equation}\label{eq:eq.p}
p_H =   \P(\chi^2(1) \geq x^2)/2^n = \left[1-\Phi(x)\right]/2^{n-1}.
\end{equation}
Likewise we can obtain the
critical value 
\begin{equation}\label{eq:alphaH.to.cH}
  c_H = \left[\Phi^{-1}(1-2^{n-1} \alpha_H) \right]^2
\end{equation}
for the test statistic \eqref{eq:eq0} or \eqref{eq:eq1}
to control the Type I error rate at some overall significance level $\alpha_H$.
Note that the overall $p$-value \eqref{eq:eq.p} cannot be larger than $1/2^n$ as it
should, since under the null hypothesis the probability to obtain $n$
positive results is $1/2^n$. We are only interested in
this case, so if at least one of the studies has a negative effect we
suggest to report the inequality $p_H > 1/2^n$, for example $p_H > 0.25$ for
$n=2$ studies.

In what follows I restrict attention to the unweighted test statistic
$X^2$ given in \eqref{eq:eq0}. Let $Z_i=z_i$ denote the observed
test statistic in the $i$-th study.  I assume that $z_i>0$ for all
$i=1,\ldots,n$, \ie all effects go in the right direction.  First note
that the smallest squared test statistic
$\zTwoMin = \min\{z_1^2, \ldots, z_n^2\}$ multiplied by the number of
studies $n$ is an upper bound on the harmonic mean
$z_H^2=n/\sum_{i=1}^n 1/z_i^2$:
\[
z_H^2 \leq n \, \zTwoMin \leq n \, z_i^2, 
\]
where the second inequality holds for all $i=1,\ldots, n$. This 
implies $x^2 \leq n^2 \, z_i^2$
for the observed test statistic $x^2$ and any study $i=1,\ldots,n$
and with equation \eqref{eq:eq.p} we obtain
\[
  \P\{\chi^2(1) \geq n^2 \, z_i^2\}/2^n \leq p_H .
\]
If $p_H \leq \alpha_H$ is required for a claim of success at level $\alpha_H$,
then obviously $\P\{\chi^2(1) \geq n^2 \, z_i^2\}/2^n  \leq \alpha_H$ must hold, 
which can be re-written as
$z_i \geq \sqrt{c_H}/{n}$ with $c_H$ given in \eqref{eq:alphaH.to.cH}. The restriction on the
corresponding $p$-values is 
\begin{equation}\label{eq:necessary}
p_i \leq 1-\Phi(\sqrt{c_H}/n). 
\end{equation}
The right-hand side of \eqref{eq:necessary}
is thus a necessary but not sufficient bound on the study-specific $p$-values for a claim of success.

It is also possible to derive the corresponding sufficient bound. Assume all
$p$-values are equal (\ie $z_1^2=\ldots=z_n^2$), then the condition
$X^2 = n\, z_i^2 \geq c_H$ implies ${z_i} \geq \sqrt{c_H}/\sqrt{n}$. 
 Note that the sufficient bound on $z_i$ differs from the
 necessary bound by a factor of $\sqrt{n}$.
The restriction on the corresponding 
$p$-values is now
\begin{equation}\label{eq:sufficient}
p_i \leq 1-\Phi(\sqrt{c_H}/\sqrt{n}).
\end{equation}
\hl{Note that for $n=1$ the necessary and sufficient bounds in
  \eqref{eq:necessary} and \eqref{eq:sufficient} both reduce to
  $\alpha_H$, as they should.}

\begin{table}[ht]
\centering
\begin{tabular}{lllllll}
  \hline
$\alpha_H$ & bound & $n=2$ & $n=3$ & $n=4$ & $n=5$ & $n=6$ \\ 
  \hline
1/1600 & necessary & 0.065 & 0.17 & 0.26 & 0.32 & 0.37 \\ 
   & sufficient & 0.016 & 0.053 & 0.099 & 0.15 & 0.20 \\ 
   \hline
1/31574 & necessary & 0.028 & 0.11 & 0.19 & 0.26 & 0.30 \\ 
   & sufficient & 0.0034 & 0.017 & 0.041 & 0.071 & 0.10 \\ 
   \hline
1/3488556 & necessary & 0.0075 & 0.058 & 0.13 & 0.19 & 0.24 \\ 
   & sufficient & 0.00029 & 0.0032 & 0.011 & 0.024 & 0.04 \\ 
   \hline
\end{tabular}
\caption{Necessary and sufficient bounds on the one-sided study-specific $p$-values for overall significance level $\alpha_H$ and different number of studies $n$} 
\label{tab:tab1}
\end{table}

\hl{The two-trials rule for drug approval is usually implemented by
requiring that each study is significant at the one-sided level
$\alpha=1/40=0.025$, so the probability of $n=2$
significant positive trials when there is no treatment effect is
$\alpha^2=1/1600 = 0.000625$. }
The necessary and sufficient bounds in \eqref{eq:necessary} and
\eqref{eq:sufficient}, respectively, are shown in Table \ref{tab:tab1} for
$\alpha_H=1/1600$
{(the two-trials rule)},
$1/31574$ (the 
four-sigma rule) and
$1/3488556$ (the 
five-sigma rule). 
\hl{The significance level of the $k$-sigma rule is based on a
  normally distributed test statistic $T \sim \Nor(0, \sigma^2)$ with zero mean 
  and defined as
$\Pr(T > k \, \sigma) = 1-\Phi(k)$. The five-sigma rule ($k=5$) was used to declare the 
discovery of the Higgs boson \citep[Section 3.2.1]{MR3127847}. The two-trials rule
corresponds to $k=3.23$, so the significance level 
of the four-sigma rule is between the two-trials rule and the five-sigma rule.}

\hl{The first line of Table \ref{tab:tab1} reveals that for level
$1/1600$}, the requirement $p_i \leq 0.065$,
$i=1,2$, is necessary for claiming success based on $n=2$ studies.  If one of the two studies
has a $p$-value larger than $0.065$, a claim of success at level
$\alpha_H=1/1600$ is thus impossible, no matter how
small the other $p$-value is. \hl{Both $p$-values being smaller than}
$0.016$ is sufficient for a claim of success
at that level.  \hl{With increasing $n$ both bounds increase, for example for 
  $n=6$ studies it is necessary that each $p$-value is smaller $0.37$
while it is sufficient that each $p$-value is smaller $0.20$.}
\hl{Decreasing the significance level from
  $1/1600$ to $1/31574$ gives similar
  bounds for $n+1$ rather than $n$ studies, and likewise for another
  decrease from $1/31574$ to 
  $1/3488556$. For example, the necessary bound is 0.17
for $\alpha_H=1/1600$ and $n=3$, 0.19
for $\alpha_H=1/31574$ and $n=4$, and again 0.19
for $\alpha_H=1/3488556$ and $n=5$.}

\subsection{\hl{Confidence intervals}}
\label{sec:CI}
 
\hl{The harmonic mean $\chi^2$ test is not directly linked to an
  overall effect estimate and a confidence interval.  However, the
  test can be inverted to obtain a confidence interval. Two extensions
  of the method are required to do so. First, we need to consider test
  statistics $Z_i = (\hat \theta_i - \mu)/\sigma_i$ for the more
  general point null hypothesis $H_0$: $\theta=\mu$.  Second, to
  compute a two-sided confidence interval we need to calculate a
  two-sided rather than one-sided $p$-value. A two-sided $p$-value
  defined as twice the one-sided $p$-value \eqref{eq:eq.p} represents
  the common scenario that an initial study is two-sided and all
  following studies aim to substantiate the effect of the first study
  including its direction, so are one-sided. The two-sided $p$-value
  $2 p_H$ can hence be evaluated not only if all effect estimates are
  positive, but also if all effect estimates are negative. If the
  effect estimates are not all in the same direction I now suggest to
  report $2 p_H>1/2^{n-1}$}.

\hl{We can now calculate a $p$-value function \citep[see][for a recent review]{Infanger2019}, 
  displaying the two-sided harmonic mean $p$-value  as a function of
  $\mu$. A two-sided confidence interval at any level $\gamma > 1- 1/2^{n-1}$ can then be
  defined as the set of $\mu$ values where the two-sided $p$-value is
  larger than $1-\gamma$. An example is given in Section
  \ref{sec:sec.app}.  
}

\section{Comparison with the two-trials rule}
\label{sec:sec3}

Suppose both studies have a positive effect in the right direction and
the observed test statistic \eqref{eq:eq0} is $X^2=x^2$. The harmonic mean
$\chi^2$
$p$-value \eqref{eq:eq.p} now reduces to
$p_H=\left[1-\Phi(x)\right]/2$.
A critical value for the test statistic
\eqref{eq:eq0} can also be calculated using \eqref{eq:alphaH.to.cH}. 
For $\alpha_H=0.025^2$ and $n=2$ we obtain
the \hl{critical} value $c_H = 9.14$.

\begin{center}
\begin{figure}[!h]
  \begin{center}
    
\begin{knitrout}
\definecolor{shadecolor}{rgb}{0.969, 0.969, 0.969}\color{fgcolor}
\includegraphics[width=\maxwidth]{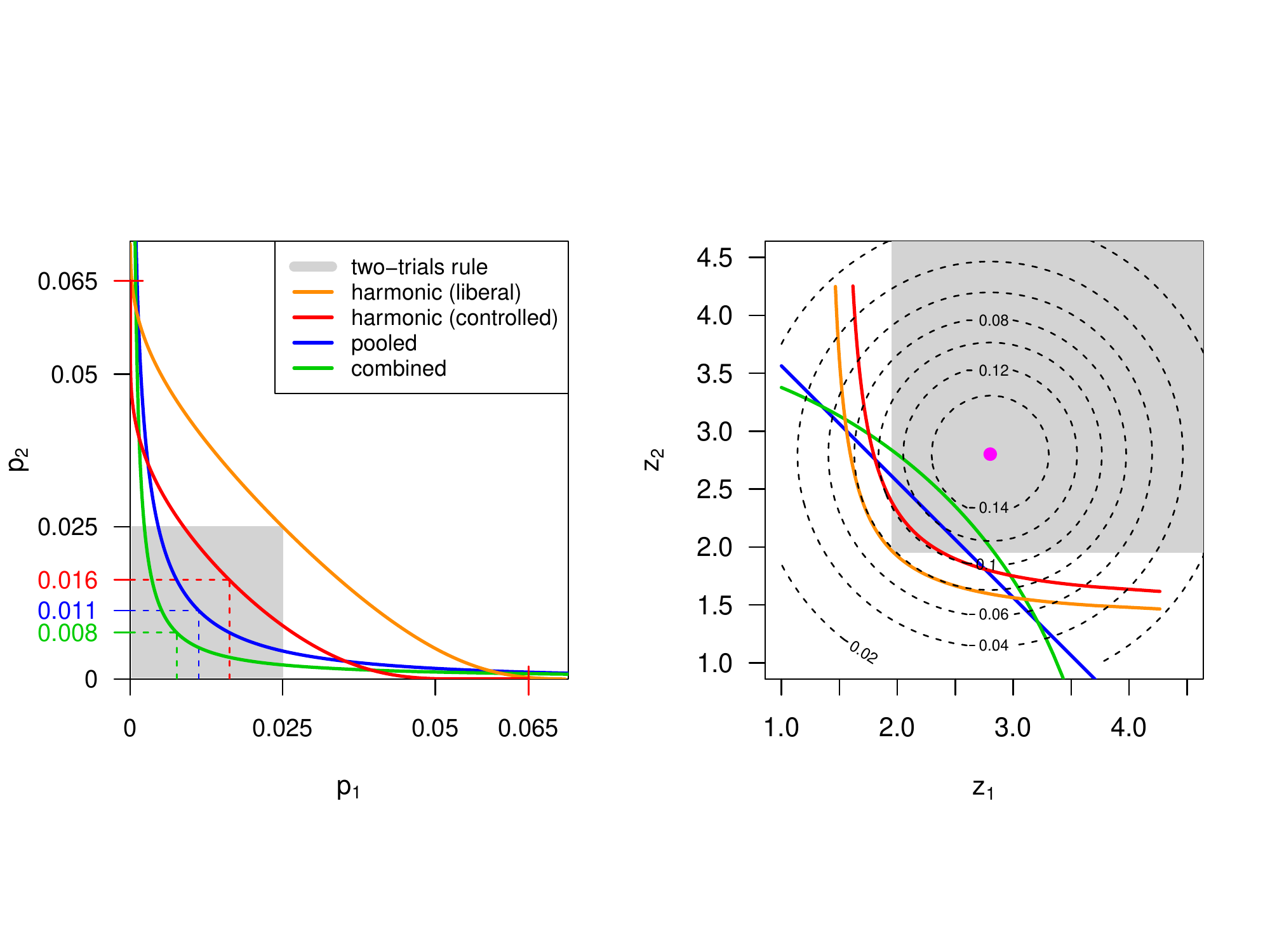} 

\end{knitrout}
\caption{Comparison of different approaches for drug approval 
  depending on the $p$-values $p_1$ and $p_2$ (left) and the 
  $Z$-values $Z_1$ and $Z_2$ (right), respectively.  The rejection
  region of the two-trials rule is shown in grey. The rejection
  regions of the other methods is below (left) or above (right) the
  corresponding curves.  All methods control the Type I error rate at
  $0.000625$ except for the liberal
  version of the harmonic mean $\chi^2$ test, which has Type I error
  rate $0.00139$.  The contour lines in the right
  plot represent the distribution of $Z_1$ and $Z_2$ under the
  alternative if the two studies have 80\% power at the
  one-sided 2.5\% significance level.}
\label{fig:fig2}
\end{center}
\end{figure}
\end{center}

Figure \ref{fig:fig2} compares the region for drug approval based on
the two-trials rule with the proposed harmonic mean $\chi^2$
test. Shown are two versions of the latter, the ``controlled'' version
based on $\alpha_H = 0.025^2$, \ie critical value $c_H=9.14$
and a ``liberal'' version with critical value
$7.68$. The latter has been computed by equating
the right-hand side of \eqref{eq:sufficient} with 0.025 and
solving for $c_H$.  The liberal version thus ensures that approval by
the two-trials rule always leads to approval by the harmonic mean
$\chi^2$ test. The Type I error rate of the liberal version is
$0.00139$, inflated by a factor of
$2.23$ compared to the $\alpha^2=0.025^2$
level. 

Also shown in Figure \ref{fig:fig2} is the corresponding region for
drug approval of the pooled and combined method, both controlled at
Type I error $0.025^2$. Both methods compensate smaller intersections
with the two-trials rejection region with additional regions of
rejection where one of the trials shows only weak or even no evidence
for an effect.  It is interesting to see that the harmonic mean
$\chi^2$ test is closer to the two-trials rule than Stouffer's pooled
or Fisher's combined method, particularly good to see in the $z$-scale
shown in the right plot of Figure \ref{fig:fig2}. The latter two
suffer from the possibility of approval if one of the $p$-values is
very small while the other one is far away from traditional
significance. A highly significant $p$-value may actually guarantee
approval through Fisher's method, no matter how large the
$p$-value from the other study is. This is not possible for Stouffer's
method, but it may still happen that the effects from the two studies
go in different directions with the combined effect being
significant. As a consequence, the sufficient $p$-value bound, shown
in the left plot of Figure \ref{fig:fig2}, is considerably smaller for
the pooled (0.011) and combined
(0.008) method than for the 
harmonic mean $\chi^2$ test (0.016) with the same Type I error rate.  
These features make both the pooled and the combined method less suitable for drug
approval.

\begin{center}
\begin{figure}[ht]
  \begin{center}

\begin{knitrout}
\definecolor{shadecolor}{rgb}{0.969, 0.969, 0.969}\color{fgcolor}
\includegraphics[width=\maxwidth]{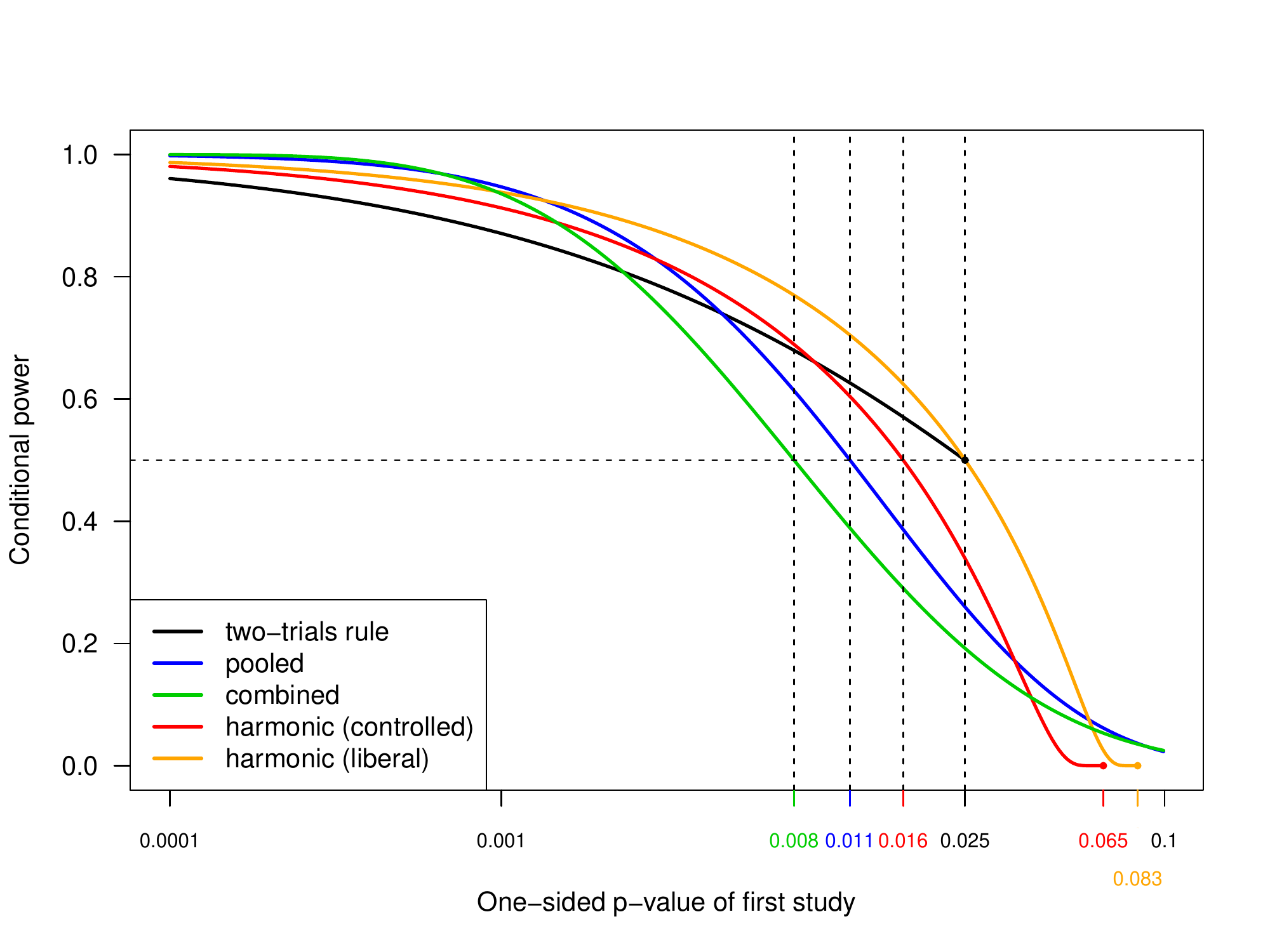} 

\end{knitrout}

\caption{Power for drug approval conditional on the one-sided $p$-value of the first study. Power values of exactly zero are omitted. } 
\label{fig:fig3}
\end{center}
\end{figure}
\end{center}

The harmonic mean $\chi^2$ test can be significant
only if both $p$-values are small ($<0.065$). 
This has been discussed in Section \ref{sec:sec2} and can also be seen
from Figure \ref{fig:fig3}, which shows the conditional power for drug
approval given the $p$-value $p_1$ from the first study. The values
represent the power to detect the observed effect from the first study
with a second study of equal design and sample size.  The two-trials
rule has conditional power as described by \cite{Goodman1992}, but with a
discontinuity at 0.025. The power curves of the two harmonic
mean $\chi^2$ tests \citep[calculated with the results given in][Section 4]{held2020} are
smooth, quickly approaching zero at $p_1 = 0.065$ respectively
$p_1 = 0.083$. Both the
combined and the pooled method have longer tails with non-zero
conditional power even for a larger $p$-value of the first study. Here the conditional
power of the combined method can be derived as
$1 - \Phi[\Phi^{-1}(p_1)-\Phi^{-1}(\min\{1, c/p_1\})]$ where
$c=\P(\chi^2(4) \geq \alpha_H)$. The  conditional
power of the pooled method turns out to be 
$1-\Phi[2 \, \Phi^{-1}(p_1)- \sqrt{2}\, \Phi^{-1}(\alpha_H)]$. 

Of central interest in drug development is often the ``project power'' for a claim of success
before the two trials are conducted \citep{Maca2002}. It
is well known \citep{Matthews2006} that under the alternative that was
used to power the two trials, the distribution of $Z_1$ and $Z_2$ is
$\Nor(\mu, 1)$ where
$\mu = \Phi^{-1}(1-\alpha) + \Phi^{-1}(1-\beta)$, where $1-\beta$ is the power of
each trial. 
We can thus simulate independent $Z_1$ and $Z_2$ for $\alpha=0.025$ and different
values of the individual trial power $1-\beta$ and compute the proportion of 
results with drug approval at level $\alpha^2$. 
This is shown in Table \ref{tab:tab2} for the different methods. 

As expected, the two-trials rule gives project power equal to $(1-\beta)^2$, since
the two trials are assumed to be independent, each significant with probability
$1-\beta$. The project power of the Type I error controlled harmonic mean $\chi^2$ test is 4
to 7 percentage points larger, depending on the power of the two trials. The
project power of the combined and pooled methods are even larger
but this comes at the price that approval may be granted even if one of the two trials was not sufficiently convincing on its own.

\begin{table}[ht]
\centering
\begin{tabular}{lrrrr}
  Trial power &\multicolumn{4}{c}{Project power} \\
 \hline
  & two-trials rule & harmonic & combined & pooled \\ 
  \hline
70 & 49 & 56 & 58 & 61 \\ 
  80 & 64 & 71 & 74 & 77 \\ 
  90 & 81 & 87 & 90 & 91 \\ 
  95 & 90 & 94 & 96 & 97 \\ 
   \hline
\end{tabular}
\caption{Individual trial power and project power of different methods for drug approval (all entries in \%)} 
\label{tab:tab2}
\end{table}

\section{Application}
\label{sec:sec.app}

Two advantages of the proposed method are that it allows for weighting
and is readily applicable to the case where results from more than 2
studies are available.  Consider again the data
shown in Table \ref{tab:tab3} on the effect of Carvedilol on
mortality. Note that all
$p$-values are below the necessary success bound 0.32 
at the level of the two-trials rule, compare Table \ref{tab:tab1}.  Only the $p$-value of
study \#239 is above the sufficient bound
0.15, otherwise we could already
claim success with the unweighted harmonic mean $\chi^2$ test.

\cite{Fisher1999b} reports Fisher's combined $p$-value, which is
0.00013. Stouffer's \hl{unweighted} pooled test 
gives the $p$-value 0.00009, the weighted version gives
$p=0.00018$. For the latter the weights have
been chosen inversely proportional to the squared standard errors 
of the associated log hazard ratios also shown in Table
\ref{tab:tab3}, see Appendix \ref{app:app2} for further details. 
The harmonic mean $\chi^2$ test gives
0.00048 (unweighted) and 0.00034
(weighted), so \hl{slightly} larger values.   
Note that all these $p$-values are smaller than the \hl{threshold 0.000625 of the two-trials rule}.

\hl{I have also calculated two confidence intervals based on the
  inversion of the weighted harmonic mean $\chi^2$ test as described
  in Section \ref{sec:CI}.  The 99.875\% confidence
  interval for the hazard ratio $\theta$ goes from
  0.17 to
  0.97. The confidence
  level is selected to be compatible with the one-sided Type I error
  rate $\alpha_H=0.000625$ of the two-trials rule, as
  $1-2 \cdot 0.000625 = 0.99875$.  The more standard
  95\% confidence interval for the hazard ratio goes
  from 0.21 to
  0.74. For comparison, a
  random-effects meta-analysis 
  gives the 95\% confidence interval
  0.25 to
  0.77 (two-sided
  $p=0.004$).  A fixed-effects
  meta-analysis gives the 95\% confidence interval
  0.32 to
  0.72.  The corresponding
  two-sided $p$-value is $0.00035$. }

Suppose now that the $p$-value in study
\#223 (the largest study with the smallest standard error) 
is twice as large, \ie
0.256 rather than 0.128.  This would be
considered as unimportant by many scientists, as both $p$-values are
non-significant anyway and far away from the \hl{standard} 0.025 significance
threshold. Keeping the standard error of the log relative risk fixed,
the \hl{estimated hazard ratio} 
in this study is now 0.83
rather than 
0.72. 

This change has a noticeable effect
on the proposed method: The unweighted and weighted harmonic
mean $\chi^2$ test $p$-values increase by a factor of 2.5 and 
7.9 to 0.0012 and
0.0027, respectively, so both would now fail the $0.025^2=0.000625$
threshold for drug approval. The $p$-values  of the unweighted and weighted Stouffer's test 
increase \hl{only} by a factor of 2.3 and 
3.5 to 0.00021 and
0.00061, respectively. \hl{Both $p$-values are 
  still below the 0.000625 threshold, and this is also the case 
  for} 
Fisher's combined $p$-value, which increases
by a factor of 1.7
to 0.00022.  
This illustrates that the harmonic mean $\chi^2$ test
is more sensitive to studies with unconvincing results,
\ie relatively small effect sizes with large $p$-values.

\section{Discussion}
\label{sec:sec5}
There is considerable variation of clinical trial evidence for newly
approved therapies \citep{Downing_etal2014}. New methods are required
to provide better inferences for the assessment of pivotal trials
supporting novel therapeutic approval. The harmonic mean $\chi^2$ test
is an attractive alternative to the two-trials rule as it has more
power at the same Type I error rate and avoids the evidence paradoxes
that may occur close to the 0.025 threshold. It provides a principled
extension to substantiate research findings from more than two trials,
requesting each trial to be convincing on its own, and allows for
weights.  \hl{It is worth noting that} the proposed method is
different from the harmonic mean $p$-value
\citep{Good1958,Wilson2019}, where the null distribution
is more difficult to compute. 

The method implicitly assumes that each of the individual trials is
well-powered for realistic treatment effects. The risk that the
harmonic mean test fails increases substantially, if some of the
trials have low power. Implementation of this new method may therefore
be seen as an incentive to use sufficiently powered and properly
conducted individual studies.  \hl{Meta-analytic techniques may 
  be more suitable if some of the studies considered are underpowered
  or if there is substantial heterogeneity between studies. }

The two-trials rule is the standard for many indications, including
many neurogenerative and cardiovascular diseases.  However, approval of
treatments in areas of high medical need may not follow the two-trials
rule. An alternative approach is conditional approval based on
``adaptive pathways'' \citep{EMA2016}, where a temporary license is 
granted based on an initial positive trial. A second post-marketing
clinical trial is then often required to confirm or revoke the initial
decision \citep{Zhang_etal2019}. This setting has much in common with replication studies
that try to confirm original results in independent investigations 
\citep{held2020,Roes2020}.

\paragraph*{Acknowledgments}
I am grateful to \hl{Mathias Drton,} Karen Kafadar, Meinhard Kieser and
Martin Posch for helpful discussions and suggestions. \hl{I also
  appreciate comments by two referees on an earlier version of this
  paper.} Support by the Swiss
National Science Foundation (Project \# 189295) is gratefully
acknowledged.

\bibliographystyle{apalike}
\bibliography{antritt}

\begin{thebibliography}{}

\bibitem[Berk and Cohen, 1979]{BerkCohen1979}
Berk, R.~H. and Cohen, A. (1979).
\newblock Asymptotically optimal methods of combining tests.
\newblock {\em Journal of the American Statistical Association},
  74(368):812--814.

\bibitem[Collett, 2003]{Collett2003}
Collett, D. (2003).
\newblock {\em {M}odelling {S}urvival {D}ata in {M}edical {R}esearch}.
\newblock Chapman \& Hall, 2nd edition.

\bibitem[Cousins, 2007]{cousins2007annotated}
Cousins, R.~D. (2007).
\newblock Annotated bibliography of some papers on combining significances or
  p-values.
\newblock \url{https://arxiv.org/abs/0705.2209}.

\bibitem[Darken and Ho, 2004]{DarkenHo2004}
Darken, P.~F. and Ho, S.-Y. (2004).
\newblock A note on sample size savings with the use of a single
  well-controlled clinical trial to support the efficacy of a new drug.
\newblock {\em Pharmaceutical Statistics}, 3(1):61--63.

\bibitem[Downing et~al., 2014]{Downing_etal2014}
Downing, N.~S., Aminawung, J.~A., Shah, N.~D., Krumholz, H.~M., and Ross, J.~S.
  (2014).
\newblock Clinical trial evidence supporting {FDA} approval of novel
  therapeutic agents, 2005-2012.
\newblock {\em JAMA: The Journal of the American Medical Association},
  311(4):368--377.

\bibitem[Drton and Xiao, 2016]{DrtonXiao2016}
Drton, M. and Xiao, H. (2016).
\newblock Wald tests of singular hypotheses.
\newblock {\em Bernoulli}, 22:38--59.

\bibitem[{European~Medical~Agency}, 2016]{EMA2016}
{European~Medical~Agency} (2016).
\newblock Adaptive pathways workshop - {Report on a meeting with stakeholders
  held at EMA on Thursday 8 December 2016}.
\newblock
  \url{https://www.ema.europa.eu/en/documents/report/adaptive-pathways-workshop-report-meeting-stakeholders-8-december-2016_en.pdf}.

\bibitem[FDA, 1998]{FDA1998}
FDA (1998).
\newblock Providing clinical evidence of effectiveness for human drug and
  biological products.
\newblock Technical report, {US\,\,Food\,\,and\,\,Drug\,\,Administration}.
\newblock
  \url{www.fda.gov/regulatory-information/search-fda-guidance-documents/providing-clinical-evidence-effectiveness-human-drug-and-biological-products}.

\bibitem[Fisher, 1999a]{Fisher1999b}
Fisher, L.~D. (1999a).
\newblock Carvedilol and the {Food and Drug Administration} ({FDA}) approval
  process: The {FDA} paradigm and reflections on hypothesis testing.
\newblock {\em Controlled Clinical Trials}, 20(1):16 -- 39.

\bibitem[Fisher, 1999b]{Fisher1999}
Fisher, L.~D. (1999b).
\newblock One large, well-designed, multicenter study as an alternative to the
  usual {FDA} paradigm.
\newblock {\em Drug Information Journal}, 33(1):265–271.

\bibitem[Fisher, 1958]{Fisher-1958}
Fisher, R.~A. (1958).
\newblock {\em {Statistical Methods for Research Workers}}.
\newblock Oliver \& Boyd, Edinburgh, 13th ed.(rev.) edition.

\bibitem[Good, 1955]{Good1955}
Good, I.~J. (1955).
\newblock On the weighted combination of significance tests.
\newblock {\em Journal of the Royal Statistical Society. Series B
  (Methodological)}, 17(2):264--265.

\bibitem[Good, 1958]{Good1958}
Good, I.~J. (1958).
\newblock Significance tests in parallel and in series.
\newblock {\em Journal of the American Statistical Association},
  53(284):799--813.

\bibitem[Goodman, 1992]{Goodman1992}
Goodman, S.~N. (1992).
\newblock A comment on replication, p-values and evidence.
\newblock {\em Statistics in Medicine}, 11(7):875--879.

\bibitem[Heard and Rubin-Delanchy, 2018]{HeardRubin-Dlanchy2018}
Heard, N.~A. and Rubin-Delanchy, P. (2018).
\newblock {Choosing between methods of combining $p$-values}.
\newblock {\em Biometrika}, 105(1):239--246.

\bibitem[Held, 2020]{held2020}
Held, L. (2020).
\newblock A new standard for the analysis and design of replication studies
  (with discussion).
\newblock {\em Journal of the Royal Statistical Society, Series A},
  {183}:431--469.

\bibitem[Hlavin et~al., 2016]{pmid26753552}
Hlavin, G., Koenig, F., Male, C., Posch, M., and Bauer, P. (2016).
\newblock {{E}vidence, eminence and extrapolation}.
\newblock {\em Statistics in Medicine}, 35(13):2117--2132.

\bibitem[Infanger and Schmidt-Trucksäss, 2019]{Infanger2019}
Infanger, D. and Schmidt-Trucksäss, A. (2019).
\newblock P value functions: An underused method to present research results
  and to promote quantitative reasoning.
\newblock {\em Statistics in Medicine}, 38(21):4189--4197.

\bibitem[Johnson, 2013]{MR3127847}
Johnson, V.~E. (2013).
\newblock Uniformly most powerful {B}ayesian tests.
\newblock {\em Annals of Statistics}, 41(4):1716--1741.

\bibitem[Kay, 2015]{kay:2015}
Kay, R. (2015).
\newblock {\em Statistical Thinking for Non-Statisticians in Drug Regulation}.
\newblock John Wiley \& Sons, Chichester, U.K., second edition.

\bibitem[Littell and Folks, 1973]{LittellFolks1973}
Littell, R.~C. and Folks, J.~L. (1973).
\newblock Asymptotic optimality of {F}isher's method of combining independent
  tests {II}.
\newblock {\em Journal of the American Statistical Association},
  68(341):193--194.

\bibitem[Maca et~al., 2002]{Maca2002}
Maca, J., Gallo, P., Branson, M., and Maurer, W. (2002).
\newblock Reconsidering some aspects of the two-trials paradigm.
\newblock {\em Journal of Biopharmaceutical Statistics}, 12(2):107--119.

\bibitem[Matthews, 2006]{Matthews2006}
Matthews, J.~N. (2006).
\newblock {\em {I}ntroduction to {R}andomized {C}ontrolled {C}linical
  {T}rials}.
\newblock Chapman \& Hall/CRC, second edition.

\bibitem[Nolan, 2018]{nolan:2018}
Nolan, J.~P. (2018).
\newblock {\em Stable Distributions - Models for Heavy Tailed Data}.
\newblock Birkhauser, Boston.
\newblock In progress, Chapter 1 online at
  \url{http://fs2.american.edu/jpnolan/www/stable/chap1.pdf}.

\bibitem[Pillai and Meng, 2016]{PillaiMeng2016}
Pillai, N.~S. and Meng, X.-L. (2016).
\newblock An unexpected encounter with {C}auchy and {L}évy.
\newblock {\em Annals of Statistics}, 44:2089--2097.

\bibitem[Roes, 2020]{Roes2020}
Roes, K. C.~B. (2020).
\newblock Discussion of "{A} new standard for the analysis and design of
  replication studies" by {Leonhard Held}.
\newblock {\em Journal of the Royal Statistical Society, Series A}, {183}:459.

\bibitem[Rosenkrantz, 2002]{Rosenkrantz2002}
Rosenkrantz, G. (2002).
\newblock Is it possible to claim efficacy if one of two trials is significant
  while the other just shows a trend?
\newblock {\em Drug Information Journal}, 36(1):875–879.

\bibitem[Senn, 2007]{senn:2007}
Senn, S. (2007).
\newblock {\em Statistical Issues in Drug Development}.
\newblock John Wiley \& Sons, Chichester, U.K., second edition.

\bibitem[Shun et~al., 2005]{Shun_etal2005}
Shun, Z., Chi, E., Durrleman, S., and Fisher, L. (2005).
\newblock Statistical consideration of the strategy for demonstrating clinical
  evidence of effectiveness—one larger vs two smaller pivotal studies.
\newblock {\em Statistics in Medicine}, 24(11):1619--1637.

\bibitem[Uchaikin and Zolotarev, 1999]{UchaikinZolotarev1999}
Uchaikin, V.~V. and Zolotarev, V.~M. (1999).
\newblock {\em Chance and Stability: Stable Distributions and Their
  Applications.}
\newblock Walter de Gruyter.

\bibitem[Westberg, 1985]{Westberg1985}
Westberg, M. (1985).
\newblock Combining independent statistical tests.
\newblock {\em The Statistician}, 34:287--296.

\bibitem[Wilson, 2019]{Wilson2019}
Wilson, D.~J. (2019).
\newblock The harmonic mean $p$-value for combining dependent tests.
\newblock {\em Proceedings of the National Academy of Sciences},
  116(4):1195--1200.

\bibitem[Zhang et~al., 2019]{Zhang_etal2019}
Zhang, A.~D., Puthumana, J., Downing, N.~S., Shah, N.~D., Krumholz, H., and
  Ross, J.~S. (2019).
\newblock Clinical trial evidence supporting {FDA} approval of novel
  therapeutic agents over three decades, 1995-2017: Cross-sectional analysis.
\newblock {\em medRxiv}.
\newblock \url{http://dx.doi.org/10.1101/19007047}.

\end{thebibliography}
\appendix{}
\part*{\appendixname}
\pdfbookmark[0]{\appendixname}{\appendixname} 

\section{The null distribution of the harmonic mean $\chi^2$ test statistic}
\label{app:app1}

Under the null hypothesis, $Z_i$, $i=1,\ldots,n$, is standard normal
distributed, so $Z_i^2$ is $\chi^2$ with 1 degree of freedom, \ie a
gamma $\Ga(1/2, 1/2)$ distribution.  The random variable $Y_i=1/Z_i^2$ is
therefore inverse gamma distributed, $Y_i \sim \IG(1/2, 1/2)$,
also known as the standard L\'evy distribution: $Y_i \sim \Levy(0, 1)$. 
More generally, the $\Levy(0, c)$ distribution corresponds to the 
$\IG(1/2, c/2)$ distribution and belongs to the class of 
stable distributions 
\citep[Section 2.3]{UchaikinZolotarev1999}.

Now $Z_1, \ldots, Z_n$ are assume to be independent, so 
$Y_1, \ldots, Y_n$ are also independent and we are interested in the
distribution of the sum $Y = Y_1 + \ldots + Y_n$, compare equation
\eqref{eq:eq0}. The standard L\'evy distribution is 
stable, which means that the sum of independent standard L\'evy random
variables is again a L\'evy random variable: $Y \sim \Levy(0, n^2)$,
which corresponds to a $\IG(1/2, n^2/2)$ distribution. Therefore
$1/Y = 1/\sum_{i=1}^n 1/Z_i^2$ follows a $\Ga(1/2, n^2/2)$
distribution and $X^2=n^2/Y$ in \eqref{eq:eq0} follows a $\Ga(1/2, 1/2)$, \ie a $\chi^2$
distribution with one degree of freedom.

The weighted version $Y = w_1 Y_1 + \ldots + w_n Y_n$ is also a L\'evy
random variable, $Y \sim \Levy(0, w^2)$ where
$w=\sum_{i=1}^n \sqrt{w_i}$, see \citet[Proposition
1.17]{nolan:2018}. Therefore $X_w^2=w^2/Y$  in \eqref{eq:eq1} also follows a $\chi^2$
distribution with one degree of freedom.  \hl{It is noteworthy that the $\chi^2(1)$
  distribution of $X^2$ respectively $X_w^2$ holds even under
  dependence of $Z_1, \ldots, Z_n$, as described by \citet[Conjecture
  6.2]{DrtonXiao2016} and proven by \citet[Theorem
  2.2]{PillaiMeng2016}.}

\section{Further details on the Carvedilol example}
\label{app:app2}

The data shown in Table \ref{tab:tab3} are taken from \citet[Table
1]{Fisher1999b} for the outcome mortality. The discussion in
\citet[page 17]{Fisher1999b} suggests that the $p$-values reported in
the table come from a log-rank test. The relative risks reported in
the table appear to be ``instantaneous relative risks'', \ie hazard
ratios. I have calculated the standard error of the log hazard ratios
from the limits of the 95\% confidence intervals also reported in the
table. Note that there is an apparent discrepancy between the
$p$-value and the confidence interval reported for study \#240, with the
one-sided log-rank $p$-value being just significant
($p$=0.0245) whereas the 95\% confidence interval for the
hazard ratio is from 0.04 to 1.14 and
includes the reference value 1. Leaving rounding errors aside, the
corresponding one-sided $p$-value from a Wald-test is
$p$=0.038.  This does not much affect the
harmonic mean $\chi^2$ test but the two-trials rule would obviously no
longer be fulfilled. The difference between log-rank and Wald is still
surprising, but a similar example has been reported in \citet[Example
3.3]{Collett2003}. I have decided to use the log-rank $p$-values as
reported, whereas the standard errors of log hazard ratios are only
used to weight the harmonic mean $\chi^2$ \hl{and Stouffer's test.
  Likewise, the fixed and random effects meta-analytic estimates are
  based on effect estimates calculated from the $p$-values and the log
  hazard ratio standard errors reported in Table \ref{tab:tab3}, but 
  the hazard ratios themselves are not used.} Finally note that mortality was not
the primary endpoint of the different studies, but \citet{Fisher1999b}
argues that ``it is the most important endpoint'' {and ``almost always
of primary importance to patients and their loved ones''.}

\end{document}